\documentclass{article}
\usepackage{spconf}
\usepackage{amsmath,epsfig}
\usepackage[OT1]{fontenc}
\usepackage{mathtools}
\usepackage{algorithm2e}
\usepackage{subcaption}
\usepackage{enumitem}
\usepackage[hyphens]{url}
\usepackage{float}
\usepackage{layouts}
\usepackage{hyperref}
\hypersetup{
colorlinks=true,
linkcolor=red,
anchorcolor=red,
citecolor=green,
urlcolor=black
}

\RestyleAlgo{ruled}

\expandafter\def\expandafter\normalsize\expandafter{%
  \normalsize  
  \setlength\abovedisplayskip{0pt}
  \setlength\belowdisplayskip{3pt}
  \setlength\abovedisplayshortskip{0pt}
  \setlength\belowdisplayshortskip{3pt}
  \setlength{\textfloatsep}{0.5\baselineskip plus 0.2\baselineskip minus 0.2\baselineskip}
  \setlist[itemize]{noitemsep, topsep=2pt} 
  \setlist[enumerate]{noitemsep, topsep=3pt} 
}

\let\OLDthebibliography\thebibliography
\renewcommand\thebibliography[1]{
  \OLDthebibliography{#1}
  \setlength{\parskip}{0pt}
  \setlength{\itemsep}{0pt plus 0.3ex}
}

\begin{document}\sloppy

\def\x{{\mathbf x}}
\def\L{{\cal L}}

\title{Complexity-Oriented Per-shot Video Coding Optimization}

%
\name{Hongcheng Zhong\textsuperscript{1,2} , Jun Xu\textsuperscript{1,2}, Chen Zhu\textsuperscript{1,2}, Donghui Feng\textsuperscript{1,2}, and Li Song\textsuperscript{1,2,3}$^{\ddagger}$}
\address{
\textsuperscript{1}Institute of Image Communication and Network Engineering, Shanghai Jiao Tong University\\
\textsuperscript{2}Cooperative Medianet Innovation Center, Shanghai Jiao Tong University\\ 
\textsuperscript{3}MoE Key Lab of Artificial Intelligence, AI Institute, Shanghai Jiao Tong University\\
Shanghai 200240, China\\ 
$^{\ddagger}$\{sj.hc\_Zhong, song\_li\}@sjtu.edu.cn
}


\maketitle

\begin{abstract}


Current per-shot encoding schemes aim to improve the compression efficiency by shot-level optimization. It splits a source video sequence into shots and imposes optimal sets of encoding parameters to each shot. Per-shot encoding achieved approximately 20\% bitrate savings over baseline fixed QP encoding at the expense of pre-processing complexity.
However, the adjustable parameter space of the current per-shot encoding schemes only has spatial resolution and QP/CRF, resulting in a lack of encoding flexibility. 
In this paper, we extend the per-shot encoding framework in the complexity dimension.
We believe that per-shot encoding with flexible complexity will help in deploying user-generated content.
We propose a rate-distortion-complexity optimization process for encoders and a methodology to determine the coding parameters under the constraints of complexities and bitrate ladders.
Experimental results show that our proposed method achieves complexity constraints ranging from 100\% to 3\% in a dense form compared to the slowest per-shot anchor.
With similar complexities of the per-shot scheme fixed in specific presets, our proposed method achieves BDrate gain up to -19.17\%.


\end{abstract}

\section{Introduction}
\label{sec:intro}

The recent COVID-19 pandemic has prompted even greater growth in the consumption of 
multimedia content on the Internet, 
bringing huge challenges in terms of storage, network bandwidth, and video encoding.

In the age of Internet video, HTTP Adaptive Streaming (HAS), which has been improved a lot in recent years, is the dominant form of distribution.
In HAS, the source content is encoded at multiple resolutions and/or 
quality levels, which allows client applications to switch to an appropriate 
version depending on the network bandwidth and display device limitations.
To determine the encoding parameters, traditional approaches applied a fixed bitrate ladder (a set of bitrate-resolution pairs) \cite{HLSAuthoringSpecification}
to all videos for simplicity and reliability, without considering the spatial-temporal
property of the video content.
The per-title encoding method was introduced in \cite{aaronPerTitleEncodeOptimization2017}, 
where individual titles were given optimal bitrate ladder based
on their spatial-temporal property.
Dynamic Optimizer and the convex hull video encoding framework, called per-shot encoding were proposed in \cite{katsavounidisDynamicOptimizerPerceptual2018}.
The per-shot encoding scheme further improved the compression efficiency
by shot-level optimization.
It has been shown in \cite{katsavounidisVideoCodecComparison2018} that the framework achieved
approximately 20\% bitrate savings over baseline fixed QP
(Quantization Parameter) / CRF (Constant Rate Factor) encoding for a variety of content, encoders, and quality metrics, at the cost of more increased complexity on pre-processing video shots.

Recent research on per-shot encoding focuses on fast encoding convex hull construction. Wu \cite{wuFastEncodingParameter2020} 
reported that
there was an inherent correlation between the content and its optimal encoding parameters.
The convex hull constructed in a fast configuration can be used to guide encoding in a slower configuration.
Wu \cite{wuEncodingParametersPrediction2021} further extended the work
by machine learning to predict the convex hull of a slow encoder using the convex hull of a fast encoder.
These methods use the slowest encoder configuration or the slowest encoder. 
Even though this achieves the best compression efficiency,
it also results in the lack of flexibility in terms of complexity.

Currently, the per-shot encoding scheme is mainly deployed in professionally generated content (PGC).
Improved compression efficiency helps save bandwidth cost on large-scale distribution.
Heavy computational complexity is affordable in this situation.
However, it is not practical to process all user-generated content (UGC) using the slowest encoder configuration because of its complexity. 
As UGC platforms is getting more and more popular in recent years, they are facing challenges of increasing cost (in terms of encoding, storage, and network bandwidth) even more than PGC.
Thus we need a methodology to constrain encoding complexity according to the
popularity of the content, trade-off the cost on encoding and bandwidth. 

The conventional complexity control research for video codec such as
\cite{dengSubjectiveDrivenComplexityControl2016, huangModelingAccelerationProperties2021}
usually uses statistical models to estimate the coding complexity at CTU level,
and restrain CTU partition depth to achieve the target complexity.
These methods can control the encoding complexity accurately within a certain range,
but they have several problems using in practical HAS encoding.
Firstly, they are mainly implemented on reference encoder, such as HM, 
without optimizations targeted for practical usage.
Secondly, the complexity control range is quite limited.

To address these limitations, we experimented the per-shot encoding framework with changeable
encoder presets. Complexity-oriented per-shot encoding framework is implemented.
The open source encoder x265 \cite{X265CodeRepository} and the BVI-1004K dataset \cite{afonsoSpatialResolutionAdaptation2018} are used for experiments and presentation of results.

Our key contributions can be summarized as follows:
\begin{itemize}
  \setlength{\itemsep}{0pt}
  \setlength{\parsep}{0pt}
  \setlength{\parskip}{0pt}
  \item We extend the conventional per-shot encoding framework in the complexity dimension
  by introducing preset into the parameter space.
  \item We propose a hyperbolic RDT model and perform rate-distortion-complexity analysis for per-shot encoding.
  \item Based on the RDT model proposed, we develop a method to optimize RD performance of representations in the specific complexity constraint.
\end{itemize}

\section{Problem statement}

\begin{figure} [htbp]
\setlength{\abovecaptionskip}{10pt plus 3pt minus 2pt}
    \centering
    \includegraphics[width=0.48\textwidth]{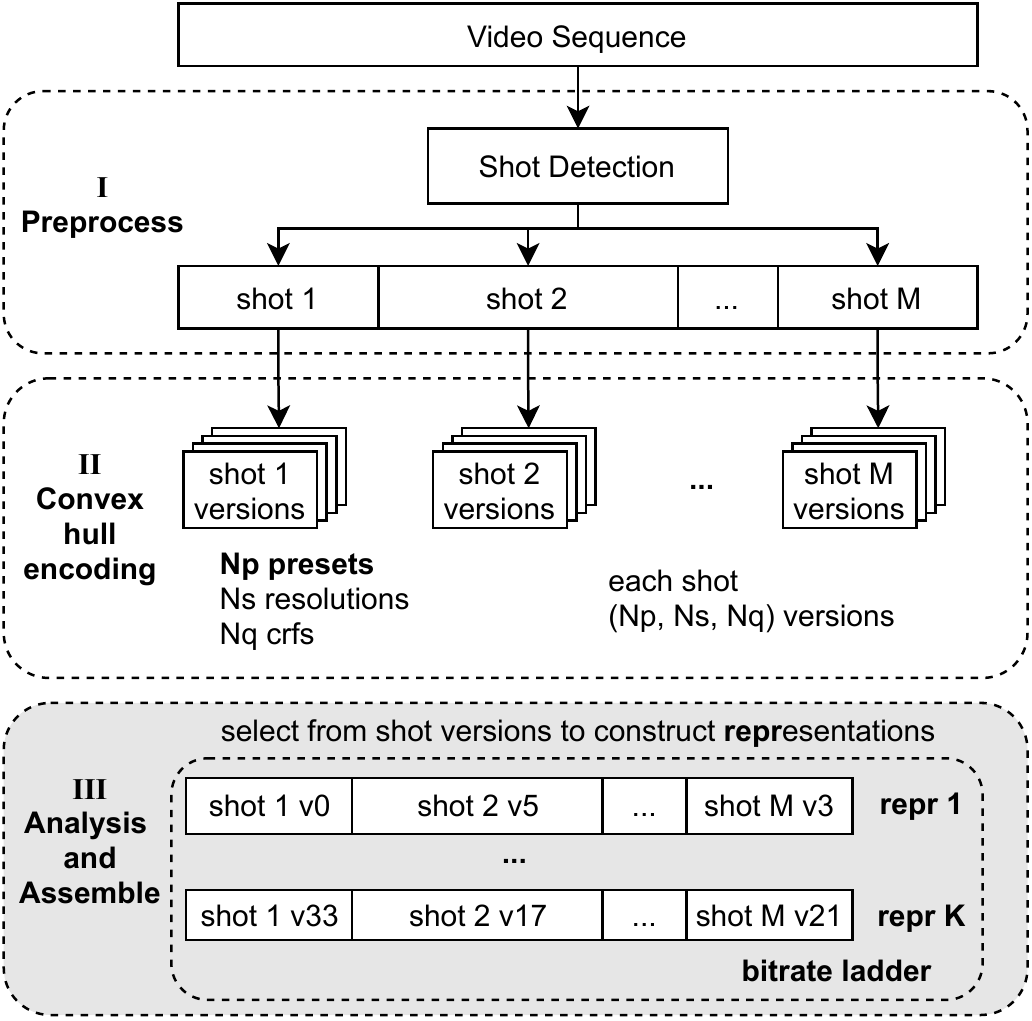}    
    \caption{Illustration of per-shot encoding workflow}
    \label{fig:flow}
\end{figure}

%
Fig.~\ref{fig:flow} shows the workflow of per-shot encoding framework: 
\begin{enumerate}[label=\Roman*.]
  \setlength{\itemsep}{0pt}
  \setlength{\parsep}{0pt}
  \setlength{\parskip}{0pt}
  \item \textbf{Preprocess}: Perform a scene-cut detection algorithm on the video sequence and split it into shots with no scene change.
  \item \textbf{Convex hull encoding}: Encode each shot with different encoding parameters, and calculate rescaled quality
  for each encoded version. 
  All encoded versions are then filtered using the convex hull algorithm, and the less efficient versions will be dropped.
  \item \textbf{Analysis and assemble}: Perform ``Dynamic Optimizer'' \cite{katsavounidisDynamicOptimizerPerceptual2018}
  on the convex hull in (II), then assemble encoded shots to a whole video sequence
  according to certain pre-specified rules. Such an assembled whole video sequence is called a \textit{representation},
  and the average bitrates of multiple representations compose a \textit{bitrate ladder}.
\end{enumerate}

\vspace{-3pt}

Since part (I) is well decoupled, scene-cut algorithms are not discussed in this paper.
The BVI-1004K dataset we used for this experiment consists of 100 sequences in 3840$\times$2160 resolution, where each sequence contains a single scene (no scene-cuts).

Practical encoders use some pre-defined configurations called \textit{preset}
to implement a flexible trade-off between encoding speed and compression efficiency.
When using slower presets, the encoder will utilize more tools and computations to achieve better compression efficiency, and vice versa.
In this paper, we consider expanding the encoding parameter space from (resolution, QP/CRF) to (preset, resolution, QP/CRF) in part (II) ``Convex hull encoding''.
More encoding parameter sets are tested for each shot in the video sequence.

An encoded version of the $i$-th shot with the parameter set $(P, S, Q)$ and the evaluated metric set $(r, d, t)$ is defined as an \textit{operating point} $OP=(P, S, Q, r, d, t)_i$:

\begin{itemize}
  \setlength{\itemsep}{0pt}
  \setlength{\parsep}{0pt}
  \setlength{\parskip}{0pt}
  \item $P$: one of $N_p$ encoder pre-defined presets.
  \item $S$: one of $N_s$ spatial resolutions.
  \item $Q$: one of $N_q$ quality levels (QP/CRF).
  \item $r$: bitrate of the encoded version.
  \item $d$: scaled distortion calculated in the original resolution (see Sec.~\ref{sec:dataset} for more details).
  \item $t$: single thread CPU time in user-space used to represent complexity.
\end{itemize}

\begin{figure}[htbp]
\captionsetup[subfigure]{aboveskip=1pt}
\begin{subfigure}[t]{.495\linewidth}
  \centering
  \includegraphics[scale=0.38]{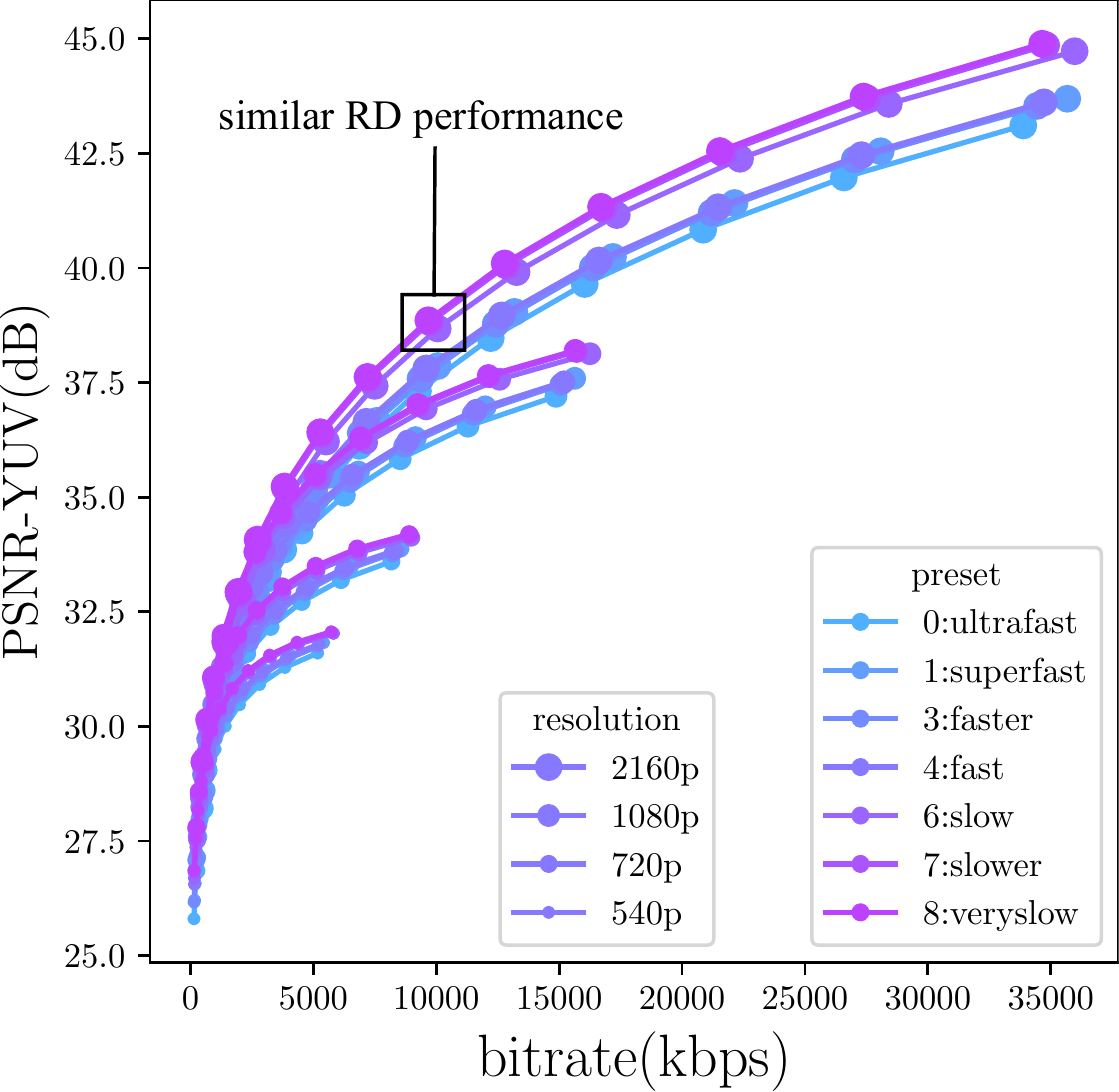}
\end{subfigure}
\hfill
\begin{subfigure}[t]{0.495\linewidth}
  \centering
  \includegraphics[scale=0.38]{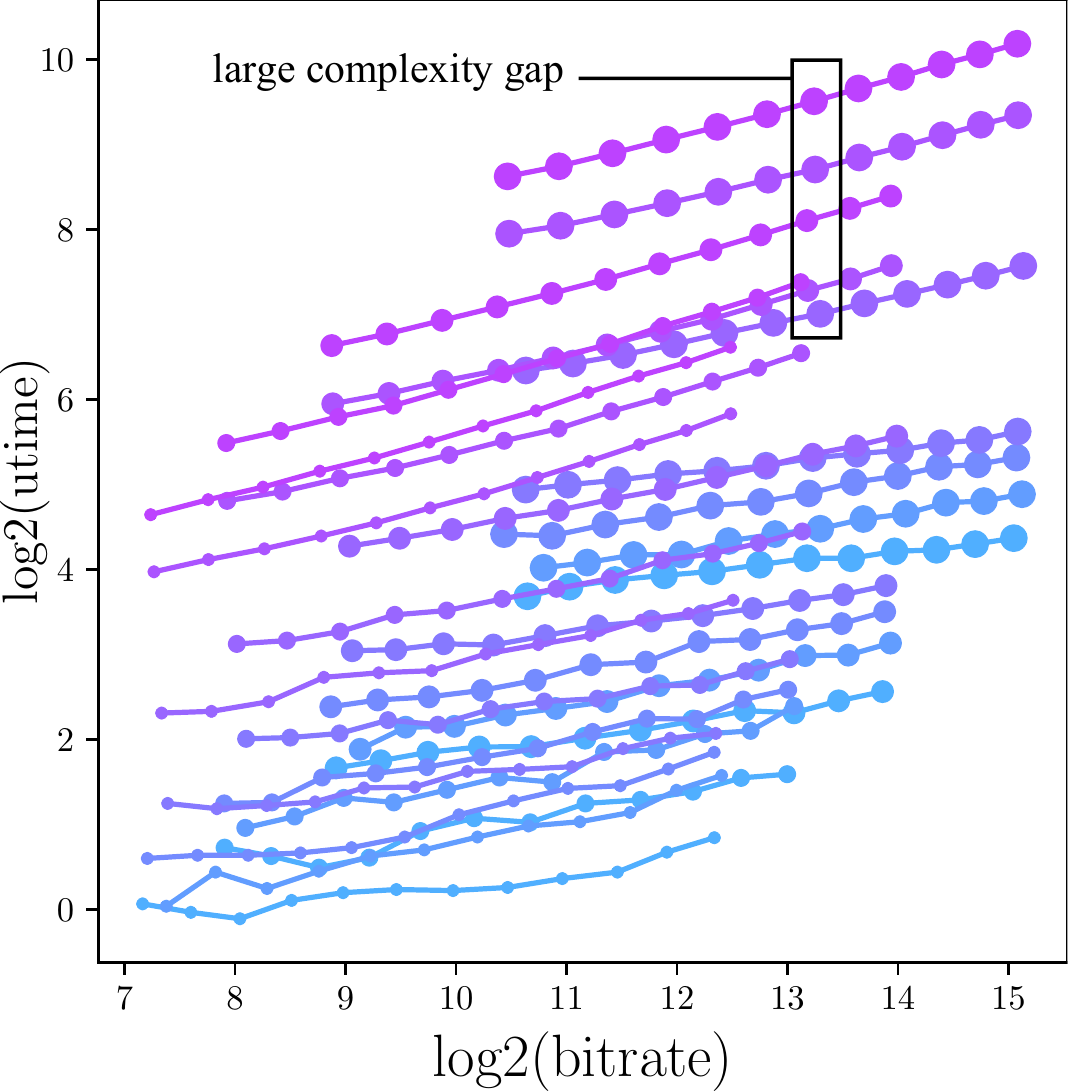}
\end{subfigure}
\vspace{-1.2\baselineskip}
\caption{Encoded versions of (preset, resolution, QP/CRF) combinations of the shot ``Aerial''. Legend with a larger size indicates larger spatial resolution, and the cooler color of the legend means the faster preset.}
\label{fig:pershot_in_presets}
\end{figure}

Fig.~\ref{fig:pershot_in_presets} shows the rate-distortion-time (R-D-T) performance of all
versions encoded in different parameters (preset, resolution, QP/CRF) of one shot.
The RD performance of encoded versions in different presets may be similar, but the complexity gap among them is quite large.
Thus Comparable RD performance can be achieved using a faster encoder preset in constrained complexity.
This motivates us to explore a complexity-oriented per-shot solution. 

Note that while part (II) ``Convex hull encoding'' is highly computational,
there is research working on the fast convex hull estimation methods \cite{wuFastEncodingParameter2020, wuEncodingParametersPrediction2021}.
Since the metric triple $(r, d, t)$ of operating points can be estimated using prediction methods, it will not be considered in this paper.

The proposed method focuses on part (III) ``Analysis and Assemble''.
As the parameter space expands to $(P,S,Q)$, rate-distortion-complexity analysis is required.
For the target bitrate and complexity level, we need a method to optimize RD performance of assembled representation.
This will be discussed in the next section.

\section{Methodology}
\subsection{Problem formulation}\label{sec:problem}


The goal of representation optimization (selecting optimal operating points in each shot to form a representation)
is to minimize the overall distortion $D$ at a given average target
bitrate $R$ and a target complexity in the form of encoding time $T$,
which is formulated by


\vspace{-6pt}
\begin{align}\label{eq:problem_v2}
  \underset{\{(P_{j}, S_{k}, Q_{l})_i\}}{\arg\min}\ &D = \sum_{i=1}^{M} d_i \\
  \text{s.t.}  \sum_{i=1}^{M} r_i &\leq R \; \text{and} \; \sum_{i=1}^{M} t_i \leq T
\end{align}
where $d_i$ , $r_i$, and $t_i$ are the distortion, bitrate, and encoding time
for operating point of the $i$-th shot, its corresponding encoding parameter set is $\{(P_{j}, S_{j}, Q_{j})_i\}$.
$M$ is the total number of shots in the sequence.
Given Lagrange multiplier $\lambda, \mu$, (\ref{eq:problem_v2}) can be
converted to an unconstrained optimization problem \cite{sullivanRatedistortionOptimizationVideo1998}

\vspace{-6pt}
\begin{equation} \label{eq:unconstrained}
  \underset{\{(P_{j}, S_{k}, Q_{l})_i\}}{\arg\min}\quad
  \sum_{i=1}^{M} \left(d_i + \lambda r_i + \mu t_i\right).
\end{equation}

Next, (\ref{eq:unconstrained}) can be solved by setting its derivative to zero:

\vspace{-3pt}
\begin{equation} \label{eq:derivation}
  \nabla (d_i + \lambda r_i + \mu t_i) = 0.
\end{equation}
Here $r_i$, $t_i$ are independent variables.
Hence solving original problem (\ref{eq:problem_v2}) is
equivalent to solving the following equation set (\ref{eq:lambda_mu_set})

\vspace{-3pt}
\begin{equation} \label{eq:lambda_mu_set}
   \left \{
    \begin{aligned}
      \frac{\partial d_i}{\partial r_i} + \lambda &= 0 \\
      \frac{\partial d_i}{\partial t_i} + \mu &= 0  \\
      \sum_{i=1}^{M} r_i &\leq R \\
      \sum_{i=1}^{M} t_i &\leq T
    \end{aligned} \right.
\end{equation}

However, it is hard to solve (\ref{eq:lambda_mu_set}) directly because of the following reasons.
\begin{enumerate}
  \setlength{\itemsep}{0pt}
  \setlength{\parsep}{0pt}
  \setlength{\parskip}{0pt}
  \item Our operating points are discrete and limited.
  \item Equation only (\ref{eq:derivation}) optimizes encoded results
  $(d_i, r_i, t_i)$ in the continuous space,
  while the corresponding encoding parameters $(P, S, Q)$ in the discrete space are actually needed.
\end{enumerate}

Therefore, we choose to solve this problem in a reverse manner.
In Sec.~\ref{sec:step1}, we propose an rate-distortion-time (RDT) model and Alg.~\ref{alg:rdt} to do R-D-T analysis and calculate $\lambda$, $\mu$ for operating points.
In Sec.~\ref{sec:step2}, we propose Alg.~\ref{alg:rdt2} to assemble
representations optimally, and generate the RDT table in the representation level.

\subsection{Convex hull analysis}\label{sec:step1}

\begin{figure}[htbp]
\captionsetup[subfigure]{aboveskip=1pt}
\begin{subfigure}[t]{.495\linewidth}
  \centering
    \includegraphics[width=\linewidth]{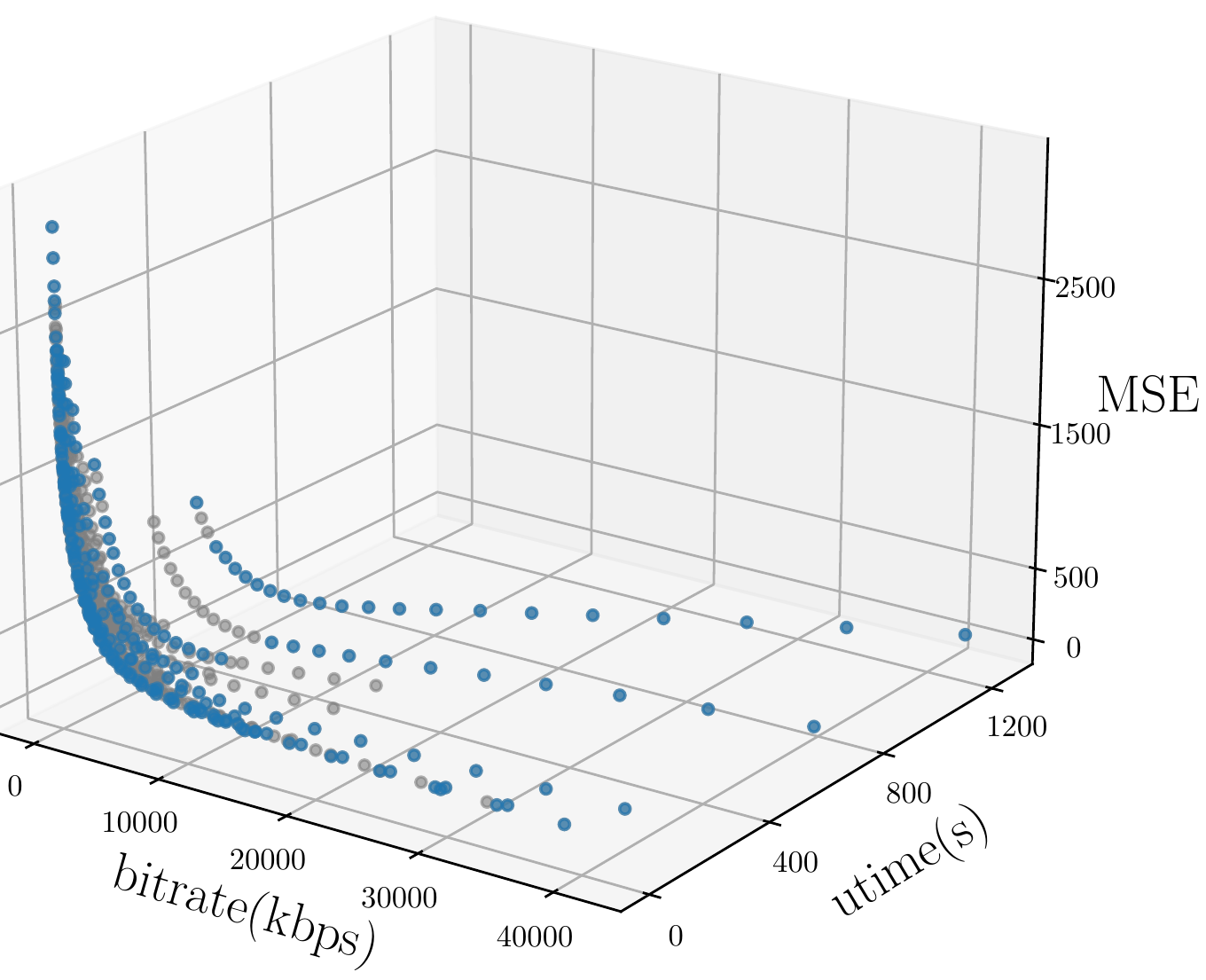}
  \caption{}
    \label{fig:convexhull3D}
\end{subfigure}
\hfill
\begin{subfigure}[t]{.495\linewidth}
    \centering
  \includegraphics[width=\linewidth]{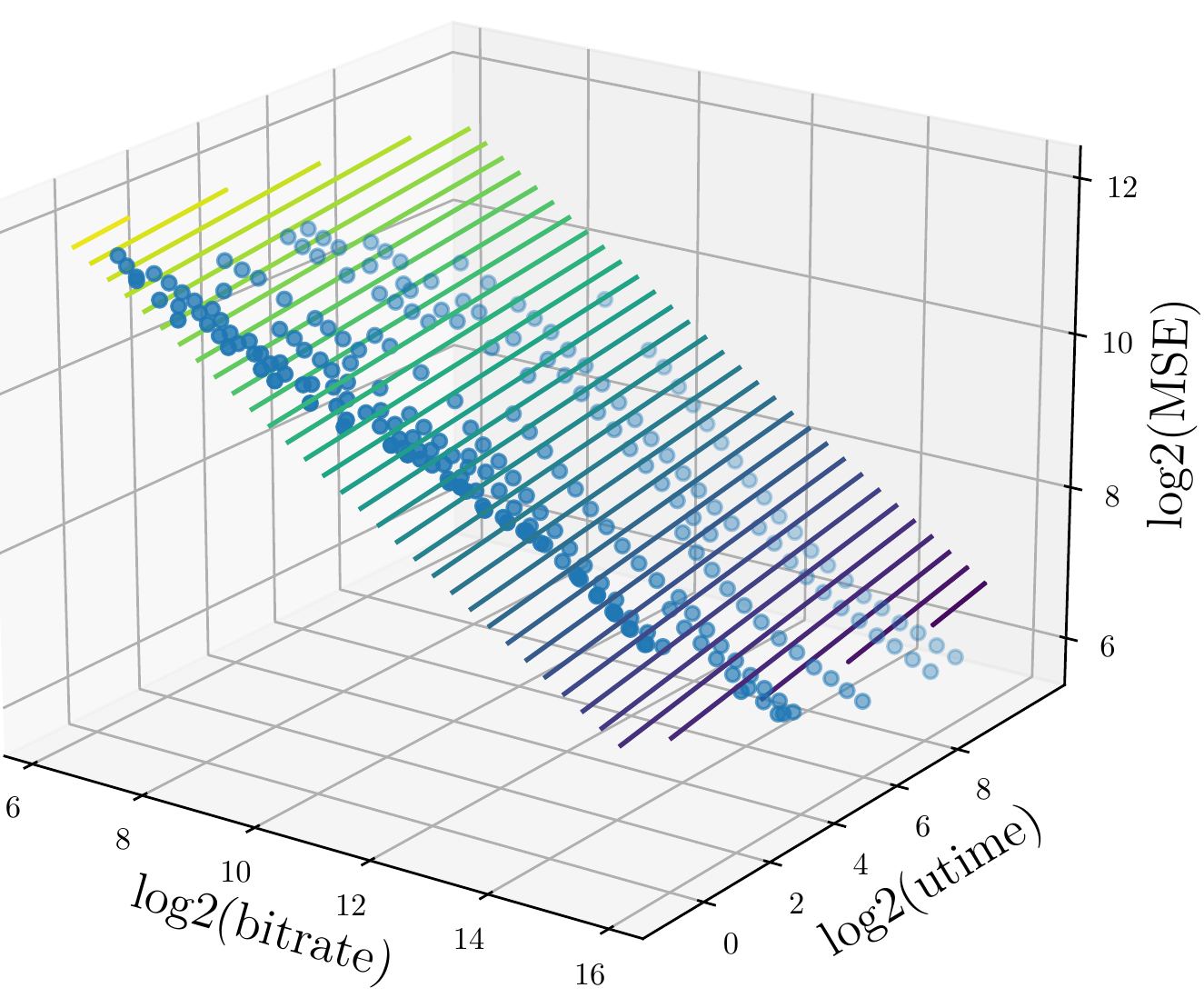}
    \caption{}
  \label{fig:fitness_aerial}
\end{subfigure}
\vfill
\begin{subfigure}[t]{\linewidth}
  \centering
  \includegraphics[width=0.5\linewidth]{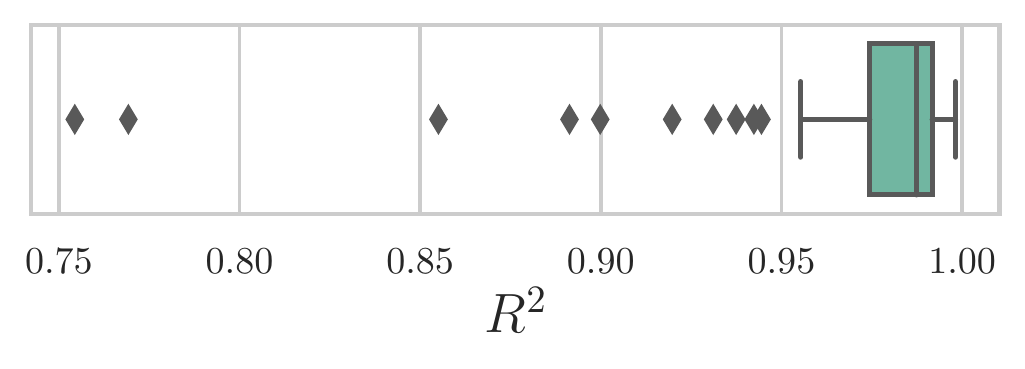}
  \caption{}
  \label{fig:rdt_fitness}
\end{subfigure}
\vspace{-0.8\baselineskip}
\caption{Fitness of hyperbolic model (\ref{eq:rdt_model}).
(a) 3D convex hull filtering, the gray points are sub-optimal operating points filtered out.
(b) An example of operating points in shot ``Aerial'', displayed in the log2 scale.
(c) The boxplot of RDT model curve fitting R-square score among 100 shots in the dataset.}
\label{fig:fitness}
\end{figure}

Alg.~\ref{alg:rdt} shows our method to process all operating points $\{(P, S, Q, r, d, t)_i\}$.
For each shot, we have $Np \times Ns \times Nq$ operating points, i.e. encoded versions.
Similar to the conventional per-shot encoding scheme, all available operating points are filtered using the convex hull method.
Encoded versions with $(r, d, t)$ metrics inside the bitrate-distortion-time plane 
are considered sub-optimal and will be filtered out as shown in Fig.~\ref{fig:convexhull3D}. 

\SetKwComment{Comment}{/* }{ */}
\begin{algorithm}[htbp]
  \caption{Rate-distortion-complexity analysis}\label{alg:rdt}
  \KwData{operating points $\{(P, S, Q, r, d, t)_i\}$}
  \KwResult{$\{\lambda\}, \{\mu\}$}
  \ForEach(){shot in shots}{
    3D convex hull filtering\;
    Curve fitting with RDT model $d = c * r ^{k_1} * t ^ {k_2}$\;
    \ForEach(){OP on convex hull}{
      Calculate $\lambda$ and $\mu$ by (\ref{eq:lambda_mu_dev})\;
    }
  }
\end{algorithm}

The RDT model (\ref{eq:rdt_model}) will work out the relationship between $d_i$ , $r_i$, and $t_i$ by RDT model fitting,

\vspace{-5pt}
\begin{equation} \label{eq:rdt_model}
  d = c \cdot r ^{k_1} \cdot t ^ {k_2}.
\end{equation}

Based on (\ref{eq:rdt_model}), (\ref{eq:lambda_mu_set}) can be rewritten as:

\vspace{-3pt}
\begin{equation} \label{eq:lambda_mu_dev}
    \begin{cases}
      \lambda_i = -\frac{\partial d_i}{\partial r_i} = -c_i k_{1i} \cdot r ^{k_{1i} - 1} \cdot t ^ {k_{2i}} \\
      \mu_i     = -\frac{\partial d_i}{\partial t_i} = -c_i k_{2i} \cdot r ^{k_{1i} } \cdot t ^ {k_{2i} - 1}
    \end{cases} 
\end{equation}

We perform curve-fitting on operating points that survived from convex hull filtering,
using average MSE as $d$, bitrate as $r$, and single thread encoding time as $t$.
Curve fitting is operated for 3-dimensional convex hull $(d_i, r_i, t_i)$ 
within each shot in the dataset. 
Details of the dataset will be introduced in Sec.\ref{sec:dataset}.


Fig.~\ref{fig:fitness_aerial} is the curve fitting result displayed in log2 scale.
Correlation coefficient $R^2$ is used to measure how well
the statistical models fit the experimental observations.
As Fig.~\ref{fig:rdt_fitness} shows,
the average of $R^2$ scores is larger than 0.974 in 100 shots.
It can be seen that from the fitted curves and the high $R^2$ scores, for the x265 encoder, the hyperbolic RDT model fits the R-D-T relationship quite well.
To verify the applicability of the model on different encoders, 
several tests have been done on the x264 \cite{X264CodeRepository}, an open source encoder for H.264/AVC.
Results show that the average $R^2$  is larger than 0.948.

With fitted hyperbolic RDT model, 
$\lambda$ and $\mu$ for all operating points can be calculated using (\ref{eq:lambda_mu_dev}).

\subsection{Optimal representation in constrained complexity}\label{sec:step2}

From Alg.~\ref{alg:rdt}, $(\lambda, \mu)$ of all operating points on convex hull for all shots are available.
Alg.~\ref{alg:rdt2} is proposed to obtain optimal representations for all
reasonable $(\lambda, \mu)$.
The RDT look-up table $(\lambda, \mu)$-representation-$(R, D, T)$ can be generated.
Therefore the desired representations can be found by checking the table with target bitrate and complexity constraint.

In Alg.~\ref{alg:rdt2}, for each $(\lambda, \mu) \in \{\lambda\}\times \{\mu\}$,
we use the RDT-cost $J = d + \lambda r + \mu t$ as criteria. 
The operating point with minimum RDT-cost in each shot will be selected.
Together, these operating points compose the representation for the given $(\lambda, \mu)$.
Finally, the average $(R, D, T)$ for the result representation can be calculated.

\begin{algorithm}[htbp]
  \caption{Representation RDT table generation}\label{alg:rdt2}
  \KwData{$\{\lambda\}\times \{\mu\}$, convex hull for each shot}
  \KwResult{RDT table: $(\lambda, \mu)$-representation-$(R, D, T)$}
  sort and concatenate all unique $\lambda$, $\mu$ \;
  \ForEach(){$\lambda$ in $\{\lambda\}$}{
    \ForEach(){$\mu$ in $\{\mu\}$}{
      \ForEach(){$shot_i$ in shots}{
        \ForEach(){$OP_j$ on convex hull}{
            RDT-cost $J_{ij} = d_{ij} + \lambda r_{ij} + \mu t_{ij}$\;
        }
        $OP_{ij} \gets  \arg\min_j \ J_{ij}$\;
      }
      $repr_{(\lambda,\mu)} \gets \{OP_{ij}\} \ i=1, 2 \dots M$\;
      calculate $(R,D,C)$ of $repr_{(\lambda,\mu)}$\;
    }
  }
\end{algorithm}

\begin{figure}[htbp]
    \captionsetup[subfigure]{aboveskip=2pt,belowskip=1pt}
    \begin{subfigure}[t]{\linewidth}
        \centering
        \includegraphics[width=0.54\linewidth]{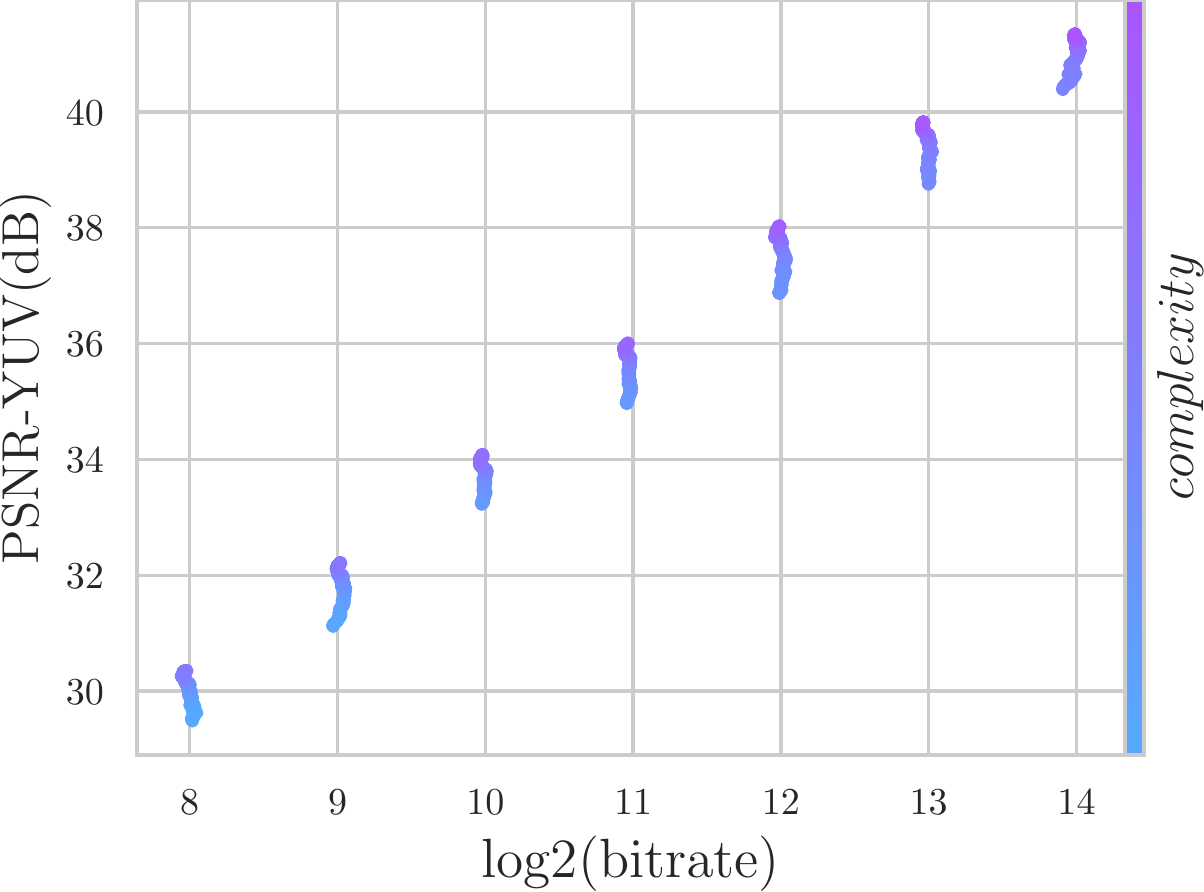}
        \caption{}
        \label{fig:complexity_lambda_mu_a}
    \end{subfigure}
        \vfill
    \begin{subfigure}[t]{.495\linewidth}
        \centering
        \includegraphics[width=\linewidth]{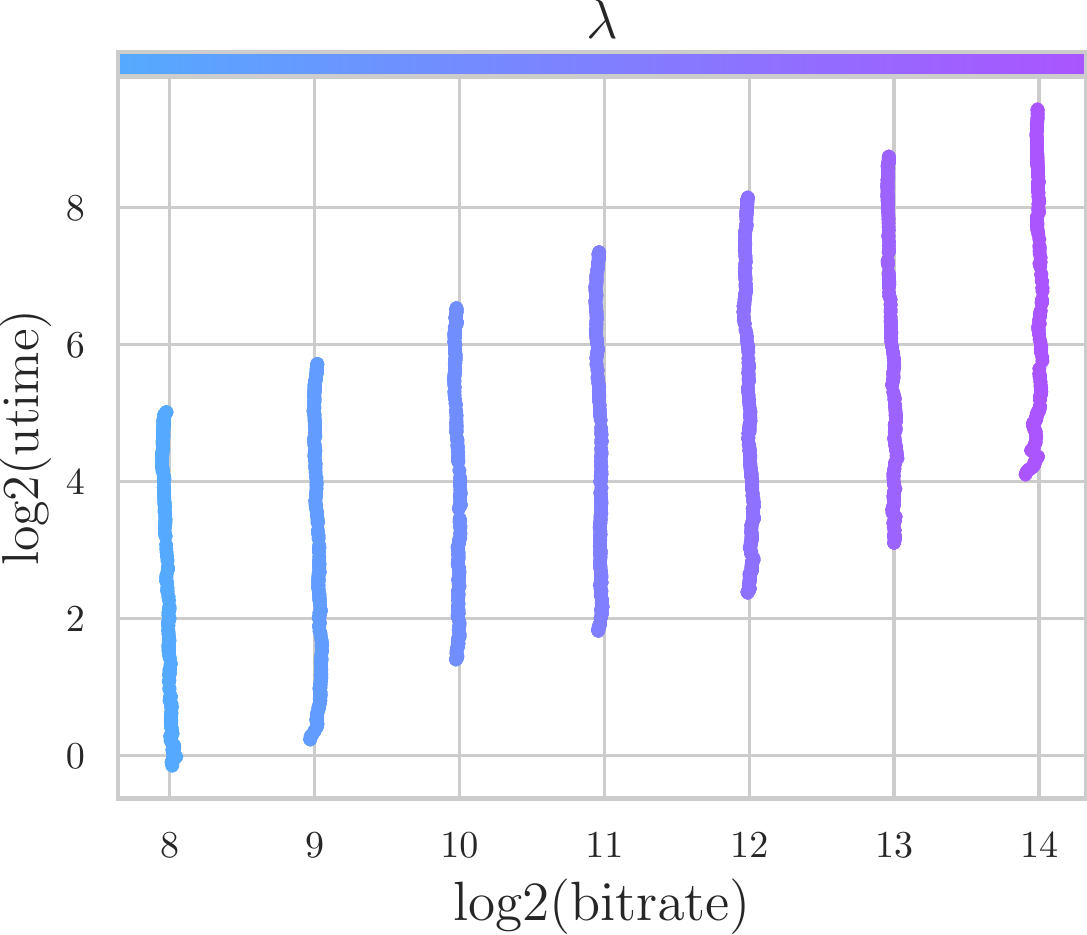}
        \caption{}
        \label{fig:complexity_lambda_mu_b}
    \end{subfigure}
        \hfill
    \begin{subfigure}[t]{0.495\linewidth}
        \centering
        \includegraphics[width=\linewidth]{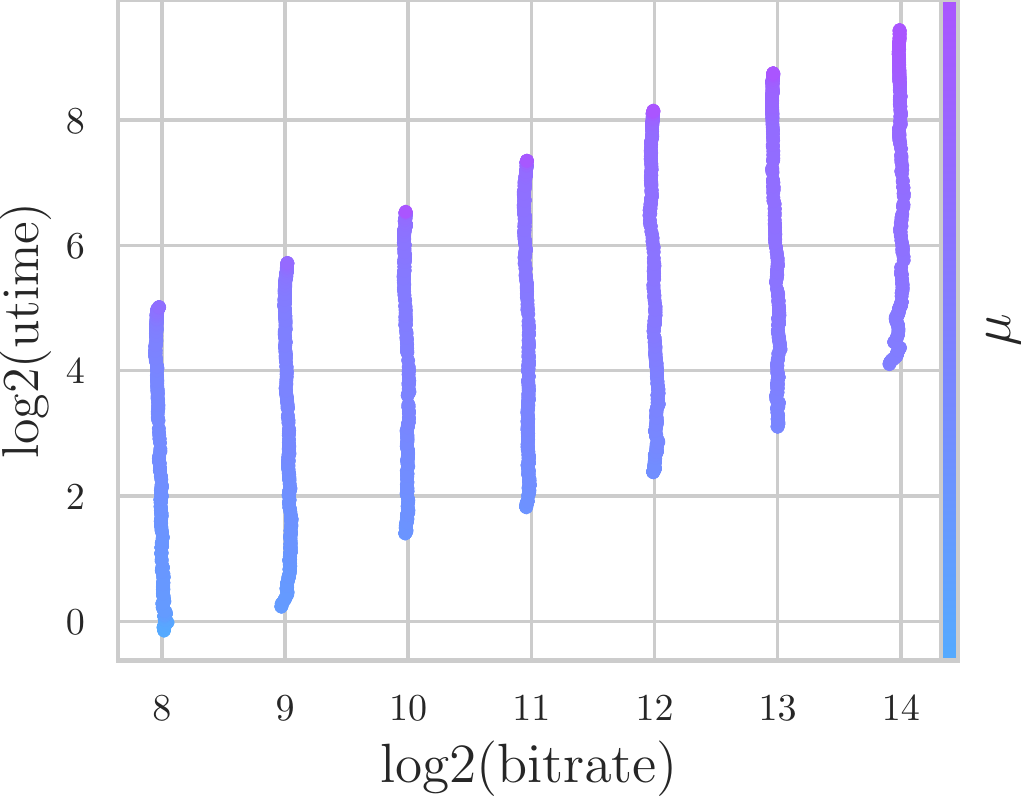}
        \caption{}
        \label{fig:complexity_lambda_mu_c}
    \end{subfigure}
\vspace{-10pt}
\caption{
Under several fixed $\lambda$ and varying $\mu$.
(a) Multiple representations for the same bitrate ladder in different complexity levels.
(b) One-to-one mapping between $\lambda$ and $R$.
(c) The complexity level gets larger when $\mu$ is increased.}
\label{fig:complexity_lambda_mu}
\end{figure}

Now we have multiple representations that achieve a similar bitrate but different in encoding complexity and compression efficiency.
Given a target bitrate, 
our methodology can provide a series of
representations with various encoding complexity ratios
compared to the slowest representation as Fig.~\ref{fig:complexity_lambda_mu_a} shows.
Then, under a specific complexity constraint, the optimal representation can be obtained from
the RDT table generated in Alg.~\ref{alg:rdt2}.
Thus the equation set (\ref{eq:lambda_mu_set}) is solved reversely.

The hyperbolic RDT model described in (\ref{eq:rdt_model}) implies that 
Lagrange multiplier $\lambda, \mu$ in (\ref{eq:lambda_mu_set}) can be used to adjust
bitrate/complexity level.
Divide two equations in (\ref{eq:lambda_mu_dev}) and let $c_i^\prime = \frac{k_{1i}}{k_{2i}}$, for the $i$-th shot, we get

\begin{equation} \label{eq:lambda_mu_2}
    \lambda_i r_i = \mu_i c_i^\prime t_i .
\end{equation}

Sum up (\ref{eq:lambda_mu_2}) for each shot $i$, 
fix $\lambda_i \approx \lambda$ and $\mu_i \approx \mu$, (\ref{eq:lambda_mu_2})
can be rewritten as:

\vspace{-3pt}
\begin{equation} \label{eq:RT_relationship}
    \lambda R = \lambda \frac{1}{M} \sum_{i=1}^M r_i 
    \approx \mu \frac{1}{M} \sum_{i=1}^M c_i^\prime t_i = \mu T^\prime ,
\end{equation}
where $T^\prime$ is the weighted encoding time that indicates complexity. 

Varying $\lambda$ will change the bitrate of the assembled representations, 
similar to the regular rate-distortion
optimization in \cite{sullivanRatedistortionOptimizationVideo1998}.
The one-to-one mappings between $\lambda$ and $R$ shown in Fig~\ref{fig:complexity_lambda_mu_b} indicate that when $\lambda$ increases, bitrate $R$ also gets higher. 
Equation (\ref{eq:RT_relationship}) indicates that for some fixed $\lambda$,
complexity level can be changing with a varying $\mu$.
This property can be used to select representations for some bitrate ladder when the complexity is constrained.
In Fig.~\ref{fig:complexity_lambda_mu_c}, for the representations with a similar bitrate,
the smaller the Lagrange multiplier $\mu$ is, the lower the complexity constraint holds.

Note that for some fixed $\lambda$, not all $\mu$ values can find new representations.
As shown in Fig.~\ref{fig:complexity_lambda_mu_c}, when target bitrate is low, large
$\mu$ can only find few corresponding representations. 
This is because
representations with a low $R$ are usually composed of shots encoded in a low resolution and/or a high QP/CRF,
resulting in much smaller complexities compared to operation points
with high resolution and low QP/CRF values. 
Likewise, on high target, a low complexity level may also be unavailable.
\vspace{-\baselineskip}

\section{Experiment Results}

\subsection{Dataset}\label{sec:dataset}
We use the BVI-1004K dataset in \cite{katsenouEfficientBitrateLadder2021}, which has 100 publicly available UHD video sequences.
The sequences have a native resolution of 3840$\times$2160,
chroma format of 4:2:0, bit depth of 10, and frame
rate of 60 fps. Each sequence contains a single scene (no scene-cuts)
including a variety of different objects/scenes/regions of
interest, camera motions, colors, and spatial activity.

In this paper, we consider 3 dimensions (preset, resolution, QP/CRF) to compose operating points.
We use the Lanczos-3 filter \cite{duchonLanczosFilteringOne1979} for spatial down/up-sampling
throughout and use rescaled quality for encoded versions in a resolution lower than 3840$\times$2160. 
The CPU used for the experiments is Intel Xeon Gold 6154 @ 3.700GHz,
and the x265 encoder version is v3.5. 
Table.~\ref{tab:dataset} shows the encoding parameters. Note that preset 2:veryfast and 5:medium are removed for continuity and monotonicity.

\vspace{-0.5\baselineskip}
\begin{table}[htbp]
\begin{center}
\caption{Encoding parameter settings} \label{tab:dataset}
\vspace{-0.5\baselineskip}
\begin{tabular}{c|p{5cm}}
  \hline
  setting & x265 encoding parameters \\
  \hline
  resolution & {2160p, 1080p, 720p, 540p} \\
  crf & $[19, 41]$, step2 \\
  preset & {0:ultrafast, 1:superfast, 3:faster, 4:fast, 6:slow, 7:slower, 8:veryslow}\\
  \hline
\end{tabular}
\end{center}
\end{table}
\vspace{-1.3\baselineskip}


The x265 encoder command template is

\noindent\texttt{\footnotesize{x265 --no-progress --input-depth 10 --input-res \{size\} --fps \{fps\} --preset \{preset\} --tune psnr --crf \{crf\} --keyint 999 --min-keyint 999 --pools 1 -F 1 --no-scenecut --no-wpp \{in\} -o \{out\}}}

We use GNU \textit{time} command to capture encoding CPU time in userspace.


\subsection{Performance comparison with the conventional per-shot}
\begin{figure}[htbp]
    \captionsetup[subfigure]{aboveskip=2pt}
    \begin{subfigure}[t]{.495\linewidth}
      \centering
      \includegraphics[width=\linewidth]{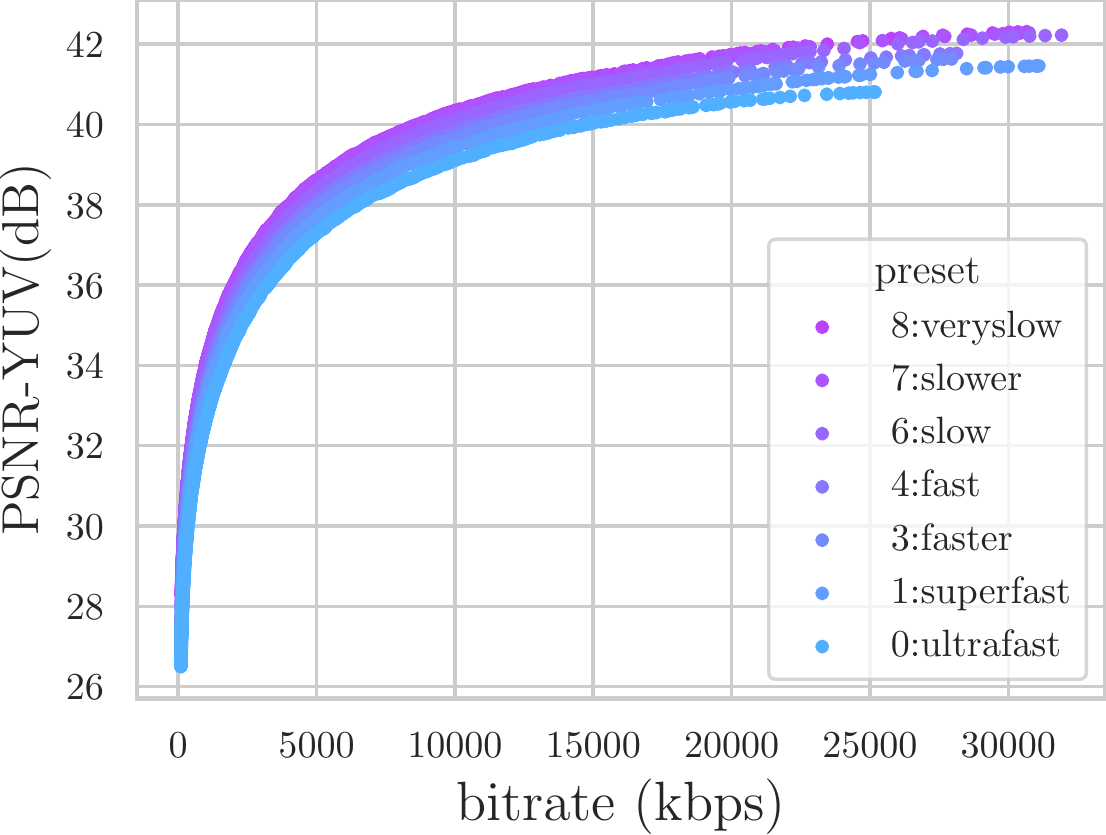}
      \caption{}
        \label{fig:normal_pershot_1}
    \end{subfigure}
    \hfill
    \begin{subfigure}[t]{.495\linewidth}
      \centering
      \includegraphics[width=\linewidth]{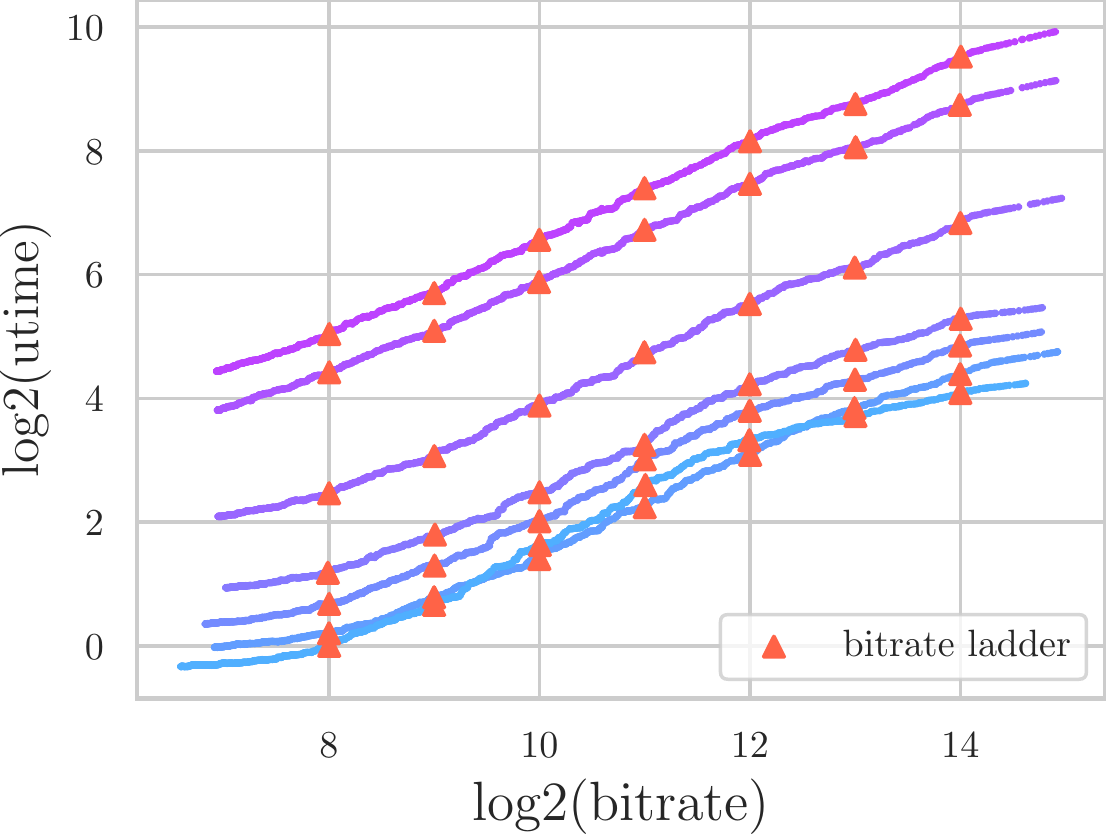}
      \caption{}
        \label{fig:normal_pershot_2}
    \end{subfigure}
    \vspace{-0.5\baselineskip}
    \caption{$(R, D, T)$ results of the conventional per-shot encoding scheme 
      restrained in specific presets,
      every point in (a) or (b) is a identical representation.
      The orange triangle markers 
      are the representations selected as the bitrate ladder for comparison.}
    \label{fig:normal_pershot}
\end{figure}

To compare with the conventional per-shot framework restrained in the specific preset,
we choose a bitrate ladder to be log2bitrate in $[8, 9, 10, 11, 12, 13, 14]$ kbps.
Fig.~\ref{fig:normal_pershot} shows the performance of conventional per-shot framework.
The curves of different presets have a similar trend: the lower the complexity chosen, the lower the PSNR in the same bitrate is achieved.

With the bitrate ladder mentioned above and the representations selected in Fig.~\ref{fig:normal_pershot_2} as the reference, 
our proposed method can find representations with an equivalent complexity to construct the bitrate ladder.


\begin{table*}[tb]
\begin{center}
\caption{BDrate compared to the conventional per-shot framework results with similar complexity.} \label{tab:performance_vs_pershot}
\begin{tabular}{c|ccccccc}
  \hline
  preset & 8:veryslow & 7:slower & 6:slow & 4:fast & 3:faster & 1:superfast & 0:ultrafast \\ \hline
  BDrate & -0.15\% & -0.28\%& -0.76\%&  -2.52\%& -3.92\%& -5.34\%& -19.17\% \\
  $r_c$ & 96.9\% &100.8\%& 101.1\%& 100.5\%& 100.0\% & 100.1\%& 102.1\% \\
  \hline
\end{tabular}
\end{center}
\vspace{-5mm}
\end{table*}

The comparison result is shown in Table.~\ref{tab:performance_vs_pershot}. 
$r_c$ is the complexity ratio of our method to the reference ones.
The result shows that
our proposed method has better BDrate performance in all presets with a similar complexity ratio.
This means the conventional per-shot framework is a subset of our method in
the extended parameter space.
The reason why presets 0:ultrafast and 8:veryslow fail to find a bitrate ladder with $r_c \approx 100\%$ 
is that some operating points are filtered out when performing convex hull filtering. 

\subsection{Performance with complexity restrained}
\begin{figure}[htbp]
\captionsetup[subfigure]{aboveskip=2pt}
\begin{subfigure}[t]{.495\linewidth}
  \centering
  \includegraphics[width=\linewidth]{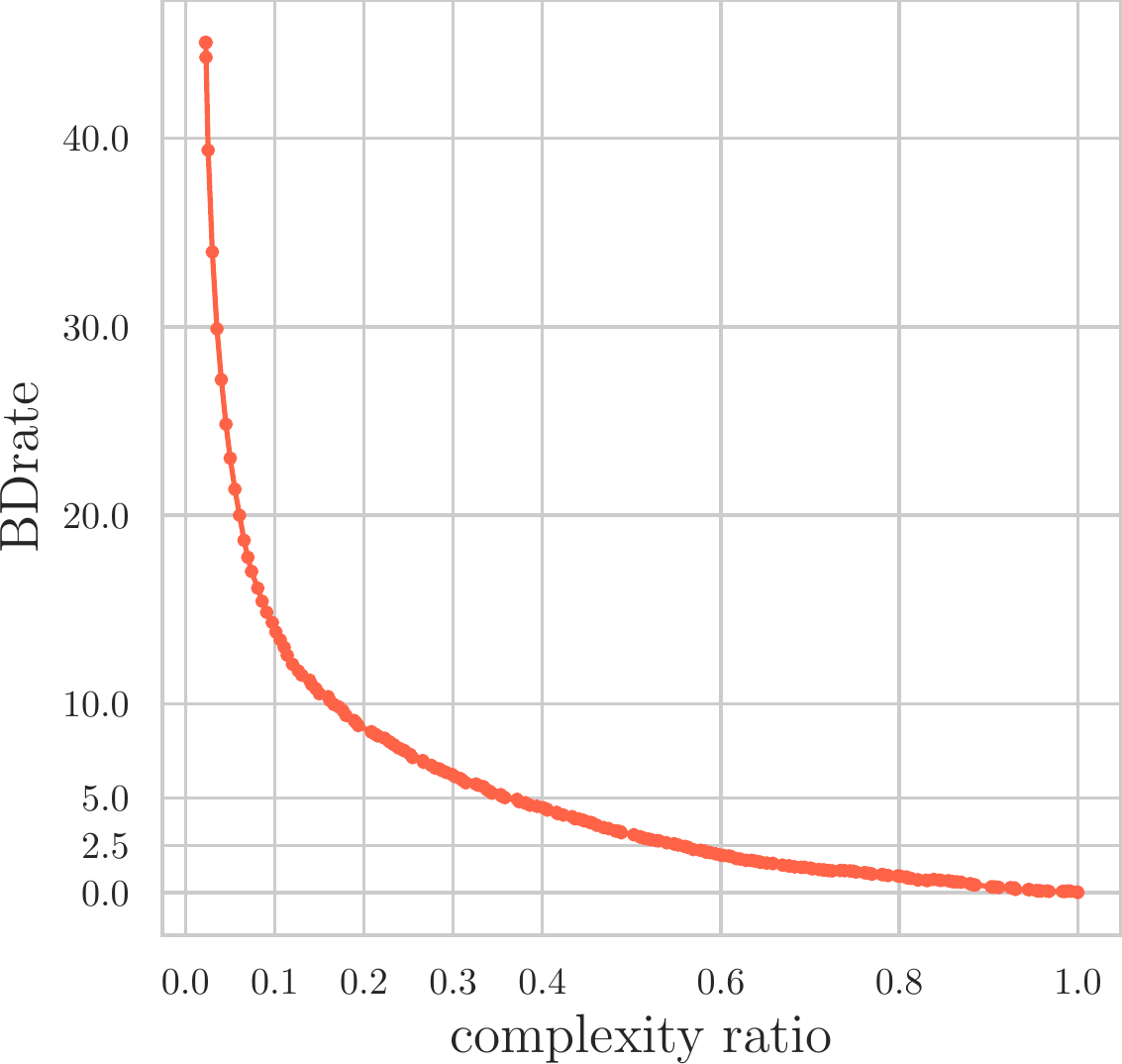}
  \caption{}
\end{subfigure}
\hfill
\begin{subfigure}[t]{0.495\linewidth}
  \centering
  \includegraphics[width=\linewidth]{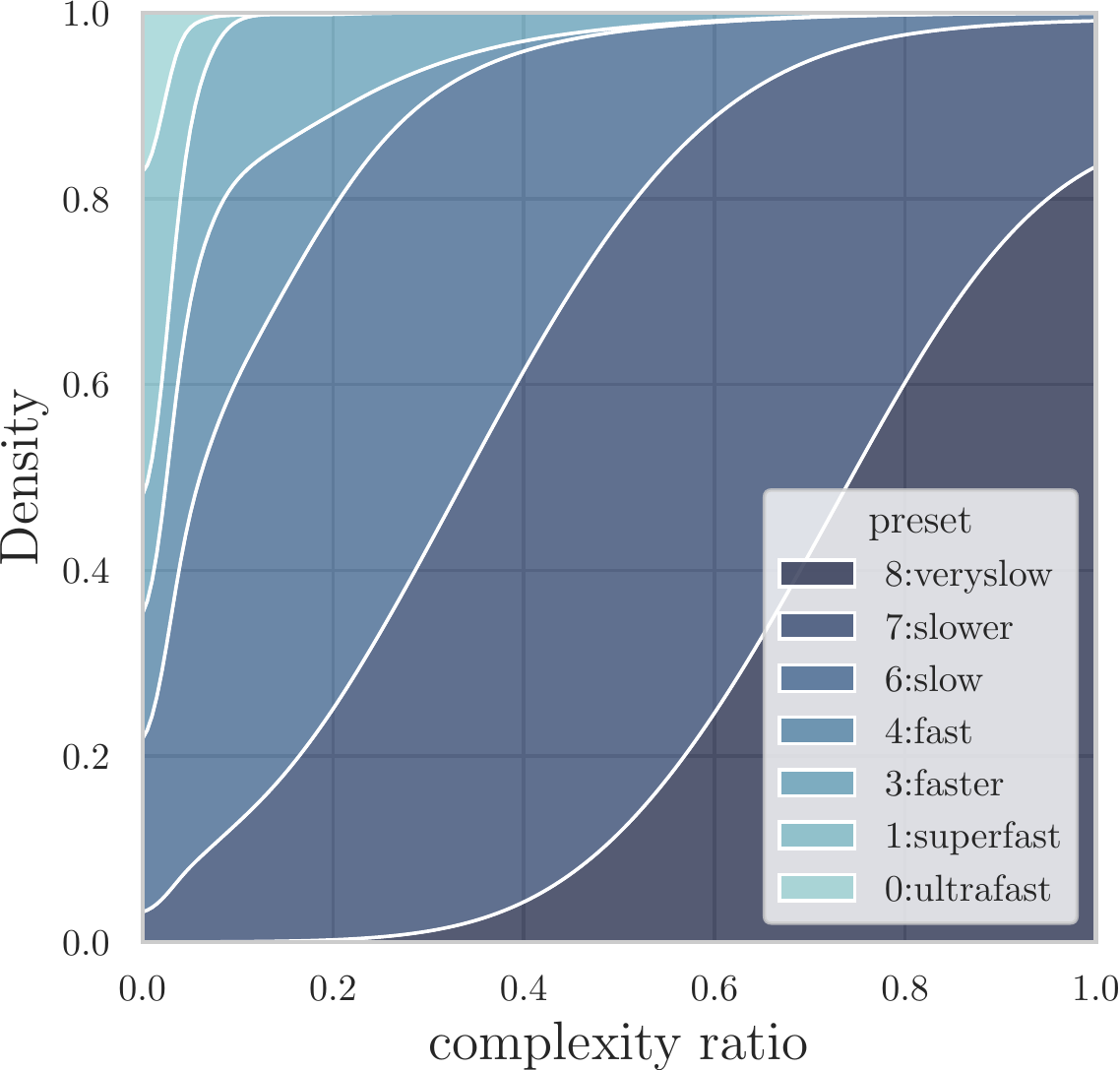}
  \caption{}
\end{subfigure}
\vspace{-0.5\baselineskip}
\caption{Performance of the complexity constrained per-shot scheme. (a) BDrate. (b) The KDEplot of preset distribution in different complexity ratios.}
\label{fig:performance}
\end{figure}

With the same bitrate ladder, our proposed method can get representations with different complexity 
constraints. The target complexity level is densely available.
Fig.~\ref{fig:performance} demonstrates the BDrate performance of representations with different
complexity levels compared to the slowest achievable representations in the bitrate ladder.
We can find that:
\begin{enumerate}
  \setlength{\itemsep}{0pt}
  \setlength{\parsep}{0pt}
  \setlength{\parskip}{0pt}
  \item Complexity constraints range from 100\% to 3\%, in which the lower
complexity leads to the more BDrate loss.
  \item Available target complexity level is dense within the control range.
  \item Encoding presets are used reasonably when the target complexity level changes.
\end{enumerate}

\section{Conclusion and Future Works}

In this paper, we extend the conventional per-shot encoding framework in the
complexity dimension. 
A hyperbolic model is designed to describe the rate-distortion-complexity relationship. With the fitted model, we propose an algorithm to assemble representations under various complexity constraints.
Experimental results show that our method allows a wide range of complexity constraints in a dense form.
With similar complexities of the per-shot scheme fixed in specific presets, our proposed method achieves decent BDrate gain.
We note that the brute convex hull encoding for all presets have about 2 times complexity of testing the slowest preset only.
Our future work aims to simplify the brute-force method by predicting the encoding results of slower presets.


\bibliographystyle{IEEEbib}
\bibliography{icme2022}

\end{document}